\documentclass[fleqn,usenatbib]{mnras}
\usepackage[T1]{fontenc}
\usepackage{graphicx}
\usepackage{amsmath}
\usepackage{amssymb}
\usepackage{txfonts}

\title[Chemical enrichment by dust grains]{Chemical enrichment of the pre-solar cloud by supernova dust grains}
\author[M. D. Goodson et al.]{
Matthew D. Goodson,$^{1}$\thanks{E-mail: mgoodson@unc.edu}
Ian Luebbers,$^{2}$
Fabian Heitsch$^{1}$
and Christopher C. Frazer$^{1}$
\\
$^{1}$Department of Physics \& Astronomy, University of North Carolina at Chapel Hill, Chapel Hill, NC 27599-3255, USA\\
$^{2}$Department of Physics \& Astronomy, Macalester College, St. Paul, MN 55105, USA\\
}
\date{Accepted 2016 July 20. Received 2016 July 18; in original form 2016 June 20}
\pubyear{2016}

\begin{document}
\label{firstpage}
\pagerange{\pageref{firstpage}--\pageref{lastpage}}
\maketitle

\begin{abstract}
The presence of short-lived radioisotopes (SLRs) in Solar system meteorites has been interpreted as evidence that the Solar system was exposed to a supernova shortly before or during its formation. Yet results from hydrodynamical models of SLR injection into the proto-solar cloud or disc suggest that gas-phase mixing may not be efficient enough to reproduce the observed abundances. As an alternative, we explore the injection of SLRs via dust grains as a way to overcome the mixing barrier. We numerically model the interaction of a supernova remnant containing SLR-rich dust grains with a nearby molecular cloud. The dust grains are subject to drag forces and both thermal and non-thermal sputtering. We confirm that the expanding gas shell stalls upon impact with the dense cloud and that gas-phase SLR injection occurs slowly due to hydrodynamical instabilities at the cloud surface. In contrast, dust grains of sufficient size ($\gtrsim 1~\micron$) decouple from the gas and penetrate into the cloud within 0.1 Myr. Once inside the cloud, the dust grains are destroyed by sputtering, releasing SLRs and rapidly enriching the dense (potentially star-forming) regions. Our results suggest that SLR transport on dust grains is a viable mechanism to explain SLR enrichment.
\end{abstract}


%%%%%%%%%%%%%%%%%%%%%%%%%%%%%%%%%%%%%%%%%%%%%%%%%%

%%%%%%%%%%%%%%%%% BODY OF PAPER %%%%%%%%%%%%%%%%%%

\section{Introduction}
\subsection{Short-lived radioisotopes}
Calcium--aluminium-rich inclusions (CAIs) in chondritic meteorites are the oldest known Solar system solids, with ages over 4.567~Gyr \citep{2002Sci...297.1678A,2010E&PSL.300..343A}. Spectroscopic analyses of CAIs reveal isotopic excesses due to the {\em in situ} decay of short-lived radioisotopes (SLRs) \citep{1977ApJ...211L.107L}, so named because of their half-lifetimes of $\lesssim$ a few Myr \citep{2001RSPTA.359.1991R,2003TrGeo...1..431M}. The radioactive decay of these SLRs, particularly $^{26}$Al, was an important source of heat during the first 10~Myr of Solar system evolution \citep{1955PNAS...41..127U}, fueling the differentiation of planetesimals \citep{2007M&PS...42.1529S} and the internal melting of ice in rocky bodies \citep{2005E&PSL.240..234T}. The sustained aqueous state due to SLRs in these bodies may have allowed the synthesis of amino acids -- the biomolecular precursors for life \citep{2014ApJ...783..140C}.

The initial abundances of some SLRs in the early Solar system (ESS) may be enhanced above the Galactic background level \citep[][however, see \citealp{2013ApJ...775L..41J}]{2006Natur.439...45D}. The presence of `live' SLRs in the ESS seems remarkable; SLRs rapidly decay and must therefore either be produced locally or quickly transported through the interstellar medium (ISM) from a nearby massive nucleosynthetic source \citep{1977ApJ...211L.107L}. In the latter case, the presence of a nearby massive star provides constraints on the birth environment of the Solar system, such as cluster size \citep{2010ARA&A..48...47A} and dynamical evolution \citep{2014MNRAS.437..946P, 2013A&A...549A..82P}. However, the conditions leading to enrichment are uncertain. The initial SLR abundances in other planet forming systems are unknown, but conditions similar to those in the ESS may be common \citep{2013ApJ...769L...8V,2013ApJ...775L..41J,2014E&PSL.392...16Y}.

The origin scenarios and initial abundances for SLRs are still a matter of debate, but it seems likely that both solar and extra-solar enrichment sources are required to explain the observed variety. Local mechanisms such as solar radiation-induced spallation reactions can produce some SLRs (e.g. $^{10}$Be) but not all (e.g. $^{60}$Fe) \citep{1976Sci...191...79H,2008ApJ...680..781G}. Although recent estimates of the initial $^{60}$Fe$/^{56}$Fe ratio argue against significant $^{60}$Fe enrichment \citep{2012E&PSL.359..248T}, the enhanced $^{26}$Al$/^{27}$Al ratio probably requires external sources \citep{2013GeCoA.110..190M}. Asymptotic giant branch (AGB) star winds \citep{1994ApJ...424..412W}, Wolf--Rayet (WR) winds \citep{1986ApJ...307..324P}, or Type II (core-collapse) supernova (SN) shock waves \citep{1977Icar...30..447C} could transport SLRs and contaminate the ESS at some phase of its evolution (e.g. pre-solar molecular cloud, pre-stellar core, or proto-planetary disc).

\subsection{Supernova enrichment}
Among the various enrichment sources, Type II supernovae (SNe) have received the most attention in the literature \citep{1977Icar...30..447C,1997ApJ...489..346F,2005ASPC..341..527O,2012ApJ...756..102P}. SNe are naturally associated with star-forming regions, and predicted SLR yields from SNe match reasonably well with ESS abundance estimates \citep{2000SSRv...92..133M}. Additional evidence is provided by the anomalous ratio of oxygen isotopes ([$^{18}$O]/[$^{17}$O]) in the Solar system, which is best explained by enrichment from Type II SNe \citep{2011ApJ...729...43Y}. 

Following the discovery of $^{26}$Al in CAIs, \citet{1977Icar...30..447C} suggested that a nearby SN could have simultaneously injected SLRs and triggered the collapse of the ESS. In this scenario, a single SN shock wave rapidly transports and deposits SLRs into an isolated marginally-stable pre-stellar core. The impinging shock wave compresses the core and triggers gravitational collapse while at the same time generating Rayleigh--Taylor (RT) instabilities at the core surface that lead to mixing of SLRs with the solar gas. \citet{1997ApJ...489..346F} first demonstrated the plausibility of this scenario with hydrodynamical simulations, and subsequent iterations of the experiment \citep{2010ApJ...708.1268B, 2012ApJ...756L...9B, 2013ApJ...770...51B, 2014ApJ...788...20B, 2015ApJ...809..103B} have defined a range of acceptable shock wave parameters (e.g. speed, width, density) for enrichment. This `triggered collapse' scenario requires nearly perfect timing and choreography. The SN must be close to the pre-stellar core ($\lesssim$ 0.1--4~pc) at the time of explosion to prevent significant SLR radioactive decay during transit; yet the SN shock must slow considerably (from $\gtrsim 2000$~km~s$^{-1}$ at ejection to $\lesssim 70$~km~s$^{-1}$ at impact) to prevent destruction of the core, requiring either large separation ($\gtrsim 10$~pc) or very dense intervening gas ($\gtrsim 100$~cm$^{-3}$). \citet{2012ApJ...745...22G} demonstrated that injection at higher velocities (up to 270~km~s$^{-1}$) may be possible, but this is yet to be confirmed in three-dimensional models.

The amount of SLRs injected in the `triggered formation' scenario is typically below observed values; both \citet{2014ApJ...788...20B} and \citet{2012ApJ...745...22G} find SLR injection efficiencies $\lesssim 0.01$, compatible with only the lowest estimates for ESS values \citep{2008ApJ...688.1382T}. Enrichment relies on hydrodynamical mixing of the ejecta into the pre-stellar gas, primarily via RT fingers \citep{2012ApJ...756L...9B}. However, the (linear) growth rates of the involved fluid instabilities depend on the square root of the density contrast \citep{1961hhs..book.....C}, resulting in an inevitable impedance mismatch between the hot, diffuse stellar ejecta and the cold, dense pre-solar core.

One possible solution to this mixing barrier problem is to concentrate the SN ejecta into dense clumps that can breach the cloud surface. The inner ejecta of Type II SNe are found to be clumpy and anisotropic in both observations \citep{2014Natur.506..339G,2015Sci...348..670B} and simulations \citep{2015A&A...577A..48W}. \citet{2012ApJ...756..102P} explore injection and mixing of clumpy SN ejecta into molecular clouds. The authors find that an over-dense clump can penetrate up to 1~pc into the target cloud, leaving a swath of enriched gas in its wake. Depending on the degree of clumpiness, the resulting enrichment can be comparable to ESS abundances.

Here, we explore an alternative mechanism to overcome the mixing barrier: the injection of SLRs via dust grains. The ejecta from both stellar winds and SNe have been predicted to condense and form dust grains \citep{1979Ap&SS..65..179C,1981ApJ...251..820E,1989ApJ...344..325K}. This prediction is supported by observations that find some SNe produce large amounts of dust ($\gtrsim 0.1~{\rm M}\odot$) soon after explosion \citep{2014ApJ...782L...2I,2015ApJ...800...50M}. In addition, meteorites contain pre-solar grains that originated in massive stars, including SNe \citep{2004ARA&A..42...39C}. Numerous authors \citep{1975ApJ...199..765C,2005ASPC..341..527O,2009ApJ...696.1854G} have suggested that these dust grains will contain SLRs, and in fact some pre-solar grains show evidence for {\em in situ} decay of $^{26}$Al \citep{2015ApJ...809...31G}. If the dust grains survive transport to the pre-solar cloud, they can dynamically decouple from the stalled shock front and penetrate into the dense gas, possibly delivering SLRs \citep{1981ApJ...251..820E,1997ApJ...489..346F}. 

\citet{2010ApJ...711..597O} have examined the role of dust grains in enrichment, considering injection into an already-formed proto-planetary disc. Although the disc's small cross-section places strong constraints on the SN distance, the authors found that over 70 per cent of dust grains with radii greater than $0.4~\micron$ can survive the passage into the inner disc where they are either stopped or destroyed. Both fates contribute SLRs to the forming star, suggesting dust grains may favorably enhance enrichment. However, injection at the disc phase may be too late; CAIs containing SLRs probably formed within the first 300,000 years of Solar system formation \citep{2005Sci...308..223Y}, prior to the proto-planetary disc phase. 

Injecting dust grains at the pre-stellar core phase may be more difficult. For grains impacting a dense pre-stellar core of number density $n \gtrsim 10^5~{\rm cm}^{-3}$, only grains with radii $a \ge 30~\micron$ are able to penetrate the stalled shock front and deposit SLRs into the core \citep{2010ApJ...717L...1B}. $30~\micron$ is greater than either simulated \citep{2015A&A...575A..95S} or meteoritic \citep{2004ARA&A..42...39C} SN grain radii (typically $a \lesssim 1~\micron$). Therefore, if injection via dust grains is to be a viable scenario, it must occur at an even earlier phase.

Enriching the pre-solar molecular cloud prior to core formation has been suggested by several authors \citep{2009ApJ...696.1854G,2009ApJ...694L...1G,2014E&PSL.392...16Y} but remains largely untested with simulations. In this scenario, one to several massive stars, possibly across multiple generations, contribute SLRs to a large star-forming region. The Solar system then forms from the enriched gas, eliminating the need for injection into a dense core. To our knowledge, the only numerical simulations  of this scenario are presented by \citet{2013ApJ...769L...8V}, with a follow-up by \citet{2016ApJ...826...22K}. The authors follow the enrichment of a massive ($\gtrsim 10^5~{\rm M}\odot$) star-forming region over 20~Myr. A turbulent periodic box is allowed to evolve subject to star formation and SN feedback. The combined effect of numerous explosions leads to an overall enrichment of $^{26}$Al and $^{60}$Fe in star-forming gas. The authors used passive particles to track SLRs, and they relied on numerical diffusion to mimic the mixing between SN ejecta and cold gas. While the resulting enrichment is broadly consistent with observed ESS values, a more detailed understanding of the injection mechanisms may be of interest.

\subsection{Motivation}
We attempt to bridge the gap between the small-scale injection scenario of \citeauthor{2010ApJ...708.1268B}, and the global, large-scale approach of \citeauthor{2013ApJ...769L...8V}, by studying the interaction of a single SN remnant with a large, clumpy molecular cloud. We focus on the details of the injection mechanism, investigating in particular the role of SLR-rich dust grains. We use hydrodynamical simulations to follow the evolution of the gas and dust over 0.3~Myr. The dust grains are decelerated by drag forces and destroyed by thermal and non-thermal sputtering, releasing SLRs into the gas phase. We estimate the amount of SLRs injected into the cloud and determine the dust grain radii needed for successful injection to occur. 

We conclude from our simulations that sufficiently large ($a \gtrsim 1~\micron$) dust grains can rapidly penetrate the cloud surface and deposit SLRs within the cloud, long before any gas can hydrodynamically mix at the cloud surface. Nearly half of all incident dust grains sputter or stop within the cloud, enriching the dense (eventually star-forming) gas. Our results suggest that dust grains offer a viable mechanism to deposit SLRs in dense star-forming gas and may be the key to reproducing the canonical Solar system SLR abundances.

We outline the numerical methods, including initial conditions and dust grain physics, in Section \ref{s:methods}. We describe measures and analytic estimates for the injection efficiency in Section \ref{s:estimates}. We present the results of our simulations in Section \ref{s:results} and discuss the implications for enrichment scenarios in Section \ref{s:discussion}. Finally, we summarize our conclusions in Section \ref{s:conclusions}.

\section{Methods}\label{s:methods}
We use a modified version of \textsc{Athena} \citep{2008ApJS..178..137S} version 4.2 to solve the equations of ideal, inviscid hydrodynamics including heating and cooling:
\begin{eqnarray}\label{eq:hydro}
\frac{\partial \rho}{\partial t} + \nabla \cdot (\rho {\bf u}) & = & 0 \\
\frac{\partial (\rho {\bf u})}{\partial t} + \nabla \cdot (\rho {\bf u}{\bf u} + P {\mathbfss I}) & = &  0 \\
\frac{\partial E}{\partial t} + \nabla \cdot [(E+P) {\bf u}] & = & n(\Gamma - n\Lambda)
\end{eqnarray} with the gas density $\rho$, the fluid velocity vector ${\bf u}$, the gas pressure $P$, the unit dyad ${\mathbfss I}$, the total energy density $E$
\begin{equation}\label{eq:eos}
E = \frac{P}{\gamma -1} + \frac{1}{2} \rho |{\bf u}|^2,
\end{equation}the number density $n \equiv \rho/(\mu m_{\rm H})$, the mass of hydrogen $m_{\rm H}$, a mean atomic weight $\mu = 1$, a heating rate $\Gamma$, and a volumetric cooling rate $\Lambda$. We also evolve several passive tracer fields:
\begin{eqnarray}
\frac{\partial \rho C_{\rm c}}{\partial t} + \nabla \cdot (\rho C_{\rm c} \bf{u}) & = & 0 \\
\frac{\partial \rho C_{\rm s}}{\partial t} + \nabla \cdot (\rho C_{\rm s} \bf{u}) & = & 0 \\
\frac{\partial \rho_{\rm d}}{\partial t} + \nabla \cdot (\rho_{\rm d} \bf{u}) & = & 0
\end{eqnarray} using colour field $C_{\rm c}$ to follow cloud material, colour field $C_{\rm s}$ to follow gas-phase SN ejecta, and four passive density fields $\rho_{\rm d}$ to track sputtered particle mass (see Section \ref{sss:sputtering}).

We use the directionally unsplit van Leer (VL) integrator \citep{2009NewA...14..139S} with second order reconstruction in the primitive variables \citep{1984JCoPh..54..174C} and the HLLC Riemann solver \citep{Toro2009}. Simulations are performed on Cartesian grids in three dimensions. We use an adiabatic equation of state with the ratio of specific heats $\gamma = C_{\rm p}/C_{\rm V} = 5/3$. Heating and cooling are included via composite curves (see Section \ref{ss:thermal}). As the cooling breaks the total energy conservation, we find it necessary to include first-order flux correction \citep{2009ApJ...691.1092L} as well as internal energy fallback (see Section \ref{ss:dualenergy}) to maintain positive states. Gravity, magnetic fields, and thermal conduction are not included. A summary of modifications made to \textsc{Athena} is given in Appendix \ref{a:mods}. 

\begin{table}
  \centering
  \caption{Summary of model parameters. Fiducial values are given in bold where necessary.}
  \label{tab:parameters}
  \begin{tabular}{llc}
    \hline
    Parameter & Definition & Values \\
    \hline
    $n_0$ & Ambient number density (cm$^{-3}$) & 1 \\
    $T_0$ & Ambient temperature (K) & 4910.58 \\
    $R_{\rm c}$ & Cloud radius (pc) & 8.8 \\
    $n_{\rm c}$ & Cloud number density (cm$^{-3}$) & $0.1 n_{\rm cl}$ = 42.33 \\
    $T_{\rm c}$ & Cloud temperature (K) & 116.02 \\
    $R_{\rm cl}$ & Clump radius (pc) & 0.05 $R_{\rm c}$ = 0.44 \\
    $n_{\rm cl}$ & Clump number density (cm$^{-3}$) & 423.25 \\
    $T_{\rm cl}$ & Clump temperature (K) & 11.60 \\
    $\phi$ & Cloud volume filling factor & 0.1, 0.3, {\bf 0.5}, 0.7, 0.9 \\
    $N_R$ & Number of cells per cloud radius & 12, 25, {\bf 50}, 100 \\
    $E_{\rm SN}$ & SN explosion energy (erg) & 10$^{51}$ \\
    $M_{\rm ej}$ & SN ejected mass (M$\odot$) & 10 \\
    $R_{\rm SNR}$ & SN remnant initial radius (pc) & 4.6 \\
    $d$ & Distance from SN centre to nearest & 17.6 \\
        & cloud edge (pc)                    &      \\ 
    $\rho_{\rm d}$ & Dust grain density (g cm$^{-3}$) & 3.0 \\
    $a$ & Dust grain radius (\micron) & 10, 1, 0.1, 0.01 \\
    $N_{\rm p}$ & Number of particles of each radius & 10$^3$, 10$^4$, {\bf 10$^5$} \\
    \hline
  \end{tabular}
\end{table}

\subsection{Setup and initial conditions}\label{ss:setup}
We initialize a spherical gas cloud in a uniform ambient medium. We use a single fluid approximation with a mean atomic weight of $\mu = 1$, treating all the gas as neutral hydrogen. The background is in thermal equilibrium with temperature $T_0 \approx 4900$~K and number density $n_0 = 1$~cm$^{-3}$, consistent with average values for the diffuse ISM \citep{1977ApJ...218..148M}. The simulation domain extends from -53 to +35 pc in $x$ and from -22 to +22 pc in $y$ and $z$. Our fiducial simulation (run F) has a resolution of $\delta_x = \delta_y = \delta_z \approx 0.17$~pc, corresponding to roughly 50 cells per cloud radius ($N_R = 50$). Table \ref{tab:parameters} summarizes our simulation parameters and values.

\subsubsection{Target molecular cloud}\label{sss:cloud}
The target molecular cloud is stationary and centred at the origin with radius $R_{\rm c} = 8.8$~pc. To approximate the substructure observed in molecular clouds, we model the cloud as a distribution of small spherical clumps of number density $n_{\rm cl} \approx 420$~cm$^{-3}$ and size $R_{\rm cl} = 0.05~R_{\rm c}$ = 0.44~pc, embedded in an intercloud medium (ICM) of number density $n_{\rm c} = 0.1~n_{\rm cl}$. The clumps are generated randomly within the cloud radius $R_{\rm c}$ up to the desired volume filling factor $\phi=0.5$. The clumps can overlap, but the density is not cumulative. The density profiles of both the cloud and the individual clumps are smoothed at the edges, and both the cloud and clumps are in pressure equilibrium with the background at $P/k_{\rm B} \approx 4900~{\rm K}~{\rm cm}^{-3}$. The clumps have a temperature $T_{\rm cl} \approx 12$~K, which also guarantees thermal equilibrium. The ICM is slightly warmer ($T_{\rm c} \approx 120$~K) and is not in strict thermal equilibrium, but the subsequent cooling is negligible and does not affect the dynamics.

The cloud edge is smoothed using the profile
\begin{equation}\label{eq:profile}
n(r) = n_0 + \frac{n_{\rm c} - n_0}{1 + (r/R_{\rm c})^{k_{\rm n}}},
\end{equation} where $r$ is the radius from the origin and $k_{\rm n}$ controls the steepness of the profile. We use $k_{\rm n}=20$ to give a steep profile. Each clump is given a similar profile by letting $n_0 \to n_{\rm c}$, $n_{\rm c} \to n_{\rm cl}$, and $R_{\rm c} \to R_{\rm cl}$. To trace cloud material, the passive colour field $C_{\rm c}$ is set to unity where $n \ge n_{\rm c}$ and zero otherwise \citep{2008ApJ...680..336S}.

\subsubsection{Supernova remnant}\label{sss:snr}
We initialize the supernova remnant (SNR) at the start of the energy-conserving phase. The shock front has expanded to $R_{\rm SNR} = (3 M_{\rm ej}/(4 \pi \rho_0))^{1/3}$ after time $t_{\rm SNR} \approx [R_{\rm SNR}(1.90E_{\rm SN}/\rho_0)^{-1/5}]^{5/2}$, where $M_{\rm ej}$ is the mass ejected from the SN and $E_{\rm SN}$ is the total energy of the SN explosion. We set $M_{\rm ej} = 10~$M$\odot$ and $E_{\rm SN} = 10^{51}$~erg, resulting in $R_{\rm SNR} \approx 4.6$~pc and $t_{\rm SNR} \approx 1000$~yr. We numerically calculate profiles for the density, radial velocity, and pressure based on the Sedov--Taylor (ST) blast-wave solutions \citep{1950RSPSA.201..159T,1959sdmm.book.....S} and interpolate these quantities on to the computational grid using a cubic spline. \textsc{Athena} uses a finite-volume method; hence if we sample only the cell-centred location (as is usually done), the resulting SNR will suffer distortion from grid effects. We find it necessary to sub-sample $8^3$ support points within each cell to construct the volume-averaged cell-centred conserved variables. 

The SNR is centred at a distance $d=2~R_{\rm c}\approx 18$~pc from the near edge of the cloud along the negative $x$-axis. This is broadly consistent with the separation distance of central stars in OB associations from bordering molecular gas, such as in Cepheus OB2 \citep{1998ApJ...507..241P}. For our target (cloud) parameters, the `radioactivity distance' \citep[equation 2, ][]{2006ApJ...652.1755L} for uniform $^{26}$Al enrichment to the initial solar system abundance is (given uncertainties in SN yield) between 10 and 20~pc. As our distance is at the upper end of this range, our enrichment estimates should be considered lower limits, as decreasing the separation would reduce the geometric dilution (see Section \ref{ss:geomdilution}).

To follow the SN gas-phase ejecta, we initialize the passive colour field $C_{\rm s}$ to unity within $R_{\rm SNR}$ and zero elsewhere. For the dust-phase ejecta, we randomly place $N_{\rm p} = 10^5$ particles of each of the four radius groups (see Section \ref{sss:sizes}) within $0.9~R_{\rm SNR}$, for a total of $4\times10^5$ particles. The particle input radius is truncated to prevent interpolation errors at the discontinuity. Particles are initialized with a radial velocity determined from the ST solution.

\subsection{Thermal physics}\label{ss:thermal}
On the time and distance scales considered here, the dynamics of the SNR should not be strongly affected by radiative cooling. \citet{1988ApJ...334..252C} and \citet{1998ApJ...500..342B} have estimated the time and location for SNR transition from the Sedov--Taylor phase to the radiative phase. For our SN parameters ($E_{\rm SN} = 10^{51}~{\rm erg}$, $n_0=1~{\rm cm}^{-3}$), the transition radius is approximately 19~pc, slightly further than the distance from the SN to the cloud surface. However, radiative cooling is expected to strongly affect the dynamics of the shock-cloud interaction. \citet{2005A&A...443..495M} have shown that cooling reduces the fragmentation and destruction of the cloud, and \citet{2008ApJ...686L.119B} find cooling by molecular species is essential to successfully inject SLR material into the pre-solar cloud. It is therefore critical to include radiative heating and cooling effects.

\begin{figure}
  \includegraphics[width=\columnwidth]{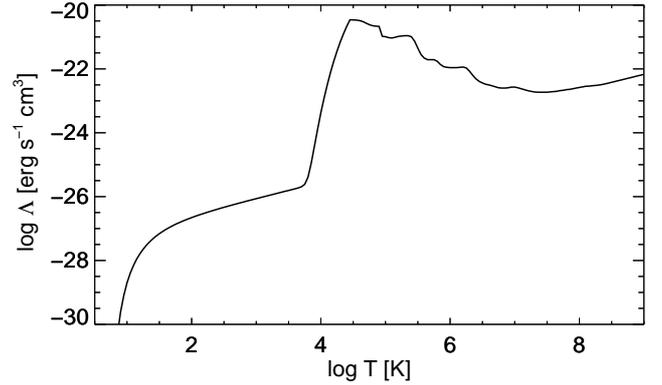}
  \caption{Volumetric cooling rate $\Lambda(T)$ (solid line) as a function of temperature $T$ from 3 to $10^{9}$~K. This composite cooling curve is constructed by blending three cooling functions from the literature: for $T < 10^4~{\rm K}$, a modified version of equation (4) from \citet{2002ApJ...564L..97K}; for $10^4~{\rm K} < T < 10^{8.5}~{\rm K}$, the C.I.E. rates from \citet{1993ApJS...88..253S}; and for $T > 10^{8.5}~{\rm K}$, the free--free rate of equation (5.15b) in \citet{1986rpa..book.....R}.}
  \label{f:coolrateplot}
\end{figure}

The temperatures in our simulation span over eight orders of magnitude, from the hot SN ejecta ($T \gtrsim 10^9$~K) to the cold molecular gas ($T \lesssim 10$~K). To cover this temperature range, we combine three standard composite cooling curves into a single cooling function, shown in Fig. \ref{f:coolrateplot}. For temperatures $T < 10^4$~K, we use a modified version of equation (4) in \citet{2002ApJ...564L..97K}:
\begin{multline}
\Lambda_{\rm KI}(T) = 2 \times 10^{-26}~\{10^7 \exp{\frac{-118400}{T+1000}} \\ + 0.014\sqrt{T}\exp{\frac{-22.75}{{\rm max}[1.0,(T-4.0)]}}\}~{\rm erg~s}^{-1}~{\rm cm}^{3}.
\end{multline}This is a fit to the cooling rates of \citet{1995ApJ...443..152W}. For temperatures $10^4~{\rm K} < T < 10^{8.5}~{\rm K}$, we use the collisional ionization equilibrium cooling rates for solar metallicity given in table 6 of \citet{1993ApJS...88..253S}. For temperatures $T > 10^{8.5}$ K, we use the free--free cooling rate given by equation (5.15b) in \citet{1986rpa..book.....R}:
\begin{equation}
\Lambda_{\rm RL}(T) = 1.42554\times10^{-27}~g~\sqrt{T}~{\rm erg~s}^{-1}~{\rm cm}^{3},
\end{equation}with a Gaunt factor $g=1.5$. The transition between regimes is smoothed with a hyperbolic tangent function. For heating, we use $\Gamma(T) = 2 \times 10^{26}~{\rm erg~s}^{-1}$ below $10^4~{\rm K}$ and smoothly transition to $\Gamma(T) = 0$ above $10^4~{\rm K}$.

Heating and cooling are implemented as source terms for the total energy (and internal energy). The cooling time-scale is typically much shorter than the hydrodynamical time-step; we therefore use an iterative explicit method (adaptive Runge--Kutta--Fehlberg) to integrate the source terms in time. The update is performed each time step via operator splitting.

\subsection{Dual energy formulation}\label{ss:dualenergy}
The cooling function requires the temperature $T$, which is proportional to the internal energy density $e = P/(\gamma -1)$ via the ideal gas law. \textsc{Athena} evolves the total energy density $E$, and the internal energy is evaluated by subtracting the kinetic energy $E_{\rm kin} \equiv \rho |{\bf u}|^2 / 2$ from the total energy. In regions where the kinetic energy is a significant fraction of the total energy, the difference will be susceptible to numerical errors and the internal energy returned may be non-physical ($e < 0$). Therefore, we simultaneously solve the internal energy equation:
\begin{equation}\label{eq:eint}
\frac{\partial e}{\partial t} + \nabla \cdot (e {\bf u}) = -P~\nabla \cdot {\bf u} + n(\Gamma - n\Lambda).
\end{equation} If the internal energy is a small fraction of the total energy ($e/E \le 10^{-3}$), we revert to using $e$ rather than $E - E_{\rm kin}$. This `Dual Energy Formulation' is also used in \textsc{Enzo} \citep{2014ApJS..211...19B} and \textsc{Flash} \citep{2000ApJS..131..273F}. Further details of our implementation are given in Appendix \ref{a:dualenergy}.

\subsection{Dust grains}\label{ss:dustgrains}
Dust grains are modelled using Lagrangian tracer particles, where each simulated particle represents a collection of dust grains with similar properties and motions. Trajectories of the particles are integrated using the fully implicit method of \citet{2010ApJS..190..297B}, which we have incorporated into the VL integrator in \textsc{Athena}. In a Cartesian coordinate system, \textsc{Athena} solves an equation of motion for each particle given by
\begin{equation}
\frac{d {\bf v_i}}{dt} = -\frac{{\bf v_i} - {\bf u}}{t_{\rm stop}},
\end{equation} with ${\bf v_i}$ the velocity vector of particle $i$, ${\bf u}$ the local gas velocity vector, and $t_{\rm stop}$ the particle stopping time due to gas drag. Neglecting grain charges and assuming only pure hydrogen gas, the (collisional) drag law is given by \citep{1979ApJ...231...77D}
\begin{equation}
\frac{d {v_i}}{dt} \approx -\frac{2 \pi a^2 n k_{\rm B} T G_0(s)}{(4/3)\pi \rho_{\rm d} a^3},
\end{equation} with
\begin{equation}
G_0(s) \approx \frac{8 s}{3 \sqrt{\pi}} (1 + \frac{9\pi}{64} s^2)^{1/2}
\end{equation} and
\begin{equation}
s \equiv (\frac{m_{\rm H} {\bf v}_{\rm rel}^2}{2 k_{\rm B} T})^{1/2},
\end{equation} where $a$ is the dust grain radius, $k_{\rm B}$ is the Boltzmann constant, $T$ is the temperature of the gas, $n$ is the gas number density, $\rho_{\rm d}$ is the internal density of the dust (which we treat as constant at $\rho_{\rm d} = 3.0$~g~cm$^{-3}$), $m_{\rm H}$ is the mass of hydrogen, and ${\bf v}_{\rm rel} \equiv {\bf v_i} - {\bf u}$ is the relative velocity difference between the dust and gas. The stopping distance is evaluated as 
\begin{equation}
t_{\rm stop} = \frac{\sqrt{\pi}}{2\sqrt{2}} \frac{a \rho_{\rm d}}{n \sqrt{m_{\rm H} k_{\rm B} T}} (1 + \frac{9\pi m_{\rm H}}{128 k_{\rm B} T} {\bf v}^2_{\rm rel})^{-1/2}.
\end{equation}The gas properties ($n$, $T$, $\bf{u}$) at each particle's location are calculated from nearby grid points using a triangular-shaped cloud (TSC) interpolation scheme \citep{1988csup.book.....H}. There is no momentum feedback from the particles on the gas.

\subsubsection{Dust grain sizes}\label{sss:sizes}
The drag force and the sputtering rates depend on the dust grain radius $a$. Since the size distribution of grains formed in SN ejecta is still a matter of debate \citep{2004ARA&A..42...39C,2007MNRAS.378..973B,2007ApJ...666..955N,2015A&A...575A..95S,2015MNRAS.454.4250M}, we follow the approach of \citet{2010ApJ...711..597O} and implement an initial `distribution' of four radii: $a = $ 10, 1, 0.1, and 0.01~$\micron$. Each radius group is initialized with the same number of particles ($N_{\rm p} = 10^5$), and the sputtered mass from each radius group is tracked using a separate passive scalar field ($\rho_{\rm d}$, see Section \ref{sss:sputtering}).

\subsubsection{Sputtering}\label{sss:sputtering}
\begin{figure}
  \includegraphics[width=\columnwidth]{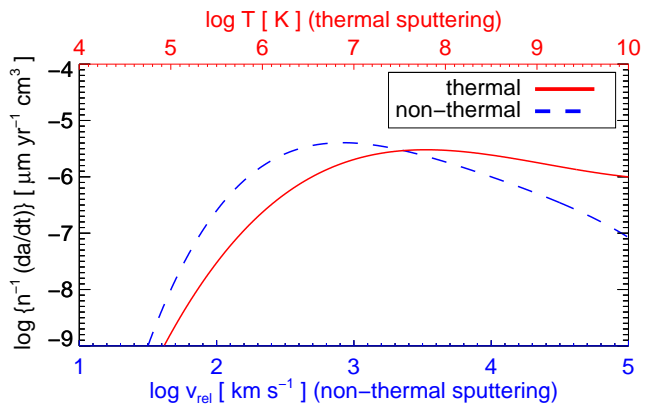}
  \caption{Polynomial fits to the thermal (solid red line) and non-thermal (dashed blue line) sputtering rates, estimated from fig. 2 of \citet{2006ApJ...648..435N}. The non-thermal sputtering varies with the relative velocity $|{\bf v}_{\rm rel}|$ between the dust and the gas (bottom axis), and the thermal sputtering varies with the gas temperature $T$ (top axis). Both rates depend on the gas number density $n$ and are given in volumetric units ($\micron~{\rm yr}^{-1}~{\rm cm}^{3}$).}
  \label{f:sputterplot}
\end{figure}

The dust grains will be eroded by both thermal and non-thermal (kinetic) sputtering. We use sputtering rates estimated from the results of \citet{2006ApJ...648..435N}, neglecting the slight differences in sputtering rate due to dust composition.

Non-thermal sputtering results from high-speed collisions of a dust grain with gas molecules and depends on the magnitude of the relative velocity $|{\bf v}_{\rm rel}|$ between the gas and the dust. For simplicity, we adopt the polynomial fit of \citet[eqs. 13,14]{2010ApJ...711..597O}\footnote{Note that \citet{2010ApJ...711..597O} contain a typographical error in the definition of $x$; cm~s$^{-1}$ should be km~s$^{-1}$.} to the non-thermal (kinetic) sputtering rates of \citet[fig. 2b]{2006ApJ...648..435N}:
\begin{multline}
y_{\rm k} = -0.1084x_{\rm k}^4 + 1.7382x_{\rm k}^3 - 10.5818x_{\rm k}^2 + 28.1292 x_{\rm k}\\ -32.7024
\end{multline} with $x_{\rm k} = {\rm log}_{10}(|{\bf v}_{\rm rel}|/1~{\rm km~s}^{-1})$ and 
\begin{equation}\label{eq:nonthermalsputtering}
(\frac{da}{dt})_{\rm k}  =-10^{y_{\rm k}}~(\frac{n}{1~{\rm cm}^{-3}})~\micron~{\rm yr}^{-1},
\end{equation}with the velocity difference between the dust and the gas $|{\bf v}_{\rm rel}|$ in km~s$^{-1}$, and the gas number density $n$. Fig. \ref{f:sputterplot} shows the volumetric non-thermal sputtering rate $n^{-1} (da/dt)$ (solid blue line) as a function of the relative velocity $|{\bf v}_{\rm rel}|$ (bottom axis).

Thermal sputtering is due to the thermal motion of the gas and depends on the temperature $T$. Similar to the procedure used by \citet{2010ApJ...711..597O} for the non-thermal sputtering rate, we generate an average fit to the thermal sputtering rates of \citet[fig. 2a]{2006ApJ...648..435N} with the polynomial
\begin{multline}
y_{\rm t} = -0.001911x_{\rm t}^4 + 0.12275x_{\rm t}^3 - 2.4011x_{\rm t}^2 + 18.6752x_{\rm t}\\ - 56.2785
\end{multline} with $x_{\rm t} = {\rm log}_{10}(T/1~{\rm K})$ and 
\begin{equation}\label{eq:thermalsputtering}
(\frac{da}{dt})_{\rm t} =-10^{y_{\rm t}}~(\frac{n}{1~{\rm cm}^{-3}})~\micron~{\rm yr}^{-1}.
\end{equation} Fig. \ref{f:sputterplot} also shows the volumetric thermal sputtering rate $n^{-1}~(da/dt)$ (dashed red line) as a function of the temperature $T$ (top axis).

We treat thermal and kinetic sputtering independently, adding the contributions to determine the erosion. However, the thermal motions of the gas will skew the relative velocity difference between the dust and the gas, particularly at high temperatures. We note that the more detailed treatment of \citet{2014A&A...570A..32B} leads to slightly lower sputtering rates in the high temperature regime, suggesting that our sputtering rates are overestimated and hence our injection efficiencies should be considered lower bounds in this regard.

The erosion rates (equations \ref{eq:nonthermalsputtering} and \ref{eq:thermalsputtering}) are applied at first order via operator splitting. A particle is assumed to be completely destroyed when its radius decreases to 1 \AA. As the particles are eroded, they release SLRs back into the gas phase. To continue tracking the sputtered SLRs in gas phase, we deposit the sputtered dust mass into a passive density field $\rho_{\rm d}$\footnote{Note that this is a passive density, rather than a colour (concentration), field. The density is a conserved quantity, whereas the concentration is not.}. This field is initially set to zero and is advected with the gas. Each initial grain radius group has its own unique passive density field. The mass is distributed into nearby cell-centred field locations using the same TSC interpolation scheme used to determine gas properties \citep{1988csup.book.....H}. Further details are given in Appendix \ref{a:sputtermass}.

\section{Enrichment estimates and measures}\label{s:estimates}
\subsection{Dust production}\label{ss:dustproduction}
We are interested in enriching a molecular cloud with SLRs from a nearby SN. The quantity of SLRs produced by a SN varies with progenitor mass \citep{2013ApJ...764...21C}, and any estimate is dominated by uncertainties in reaction rates \citep{2011ApJS..193...16I} and progenitor models \citep{2007PhR...442..269W}. Of this amount, some fraction will condense into dust grains of various sizes \citep{2015A&A...575A..95S,2015MNRAS.454.4250M}. Furthermore, the dust grains that form behind the SNR forward shock will subsequently be processed by the reverse shock  \citep{2007MNRAS.378..973B,2007ApJ...666..955N,2016A&A...589A.132B,2016A&A...587A.157B}. Calculations of dust grain processing in the reverse shock predict survival rates of 0--100 per cent, depending on the grain size, grain composition, and local gas density  \citep{2007ApJ...666..955N,2007MNRAS.378..973B,2010ApJ...715.1575S,2012ApJ...748...12S}. Additionally, inhomogeneities in the SNR produce small clumps of higher density that may shield the forming dust grains from destruction \citep{2014A&A...564A..25B,2016A&A...589A.132B,2016A&A...590A..65M}. For simplicity, we assume a homogeneous SNR and background medium. Because we begin our simulations at the end of the free-expansion phase, we neglect processing by the reverse shock. We therefore assume at least some amount of dust has survived and is still well-coupled to the gas, consistent with 1D simulations \citep{2016A&A...589A.132B,2016A&A...587A.157B}. Our calculations are normalized such that the condensation efficiency and survival rate do not affect the evolution.

\subsection{Geometric dilution}\label{ss:geomdilution}
As the SNR expands, the ejecta become distributed over a larger surface area. For a spherical target of radius $R$, at a distance $d$ from the SNR centre, the fraction of the total ejecta incident on the target cross-section is
\begin{equation}
\eta_{\rm geom} = \frac{\pi R^2}{4 \pi d^2}.
\end{equation}For our fiducial set-up, $d \approx 18$~pc and $R \approx 9$~pc; then $\eta_{\rm geom} \approx 0.06$. This factor is used to normalize our injection efficiency $\eta$ (see Section \ref{ss:injectioneff}).

\subsection{Injection efficiency}\label{ss:injectioneff}
The mixing of incident material with a target has been the subject of much previous numerical work, both in the context of the standard shock-cloud interaction \citep{1995ApJ...454..172X,2008ApJ...680..336S,2009MNRAS.394.1351P} and in Solar system enrichment \citep{2012ApJ...756L...9B, 2010ApJ...711..597O}. Defining a good measure of the mixing is difficult and depends on the context. We therefore quantify the mixing in two ways.

For the shock-cloud interaction, the mixing fraction is typically defined by the dilution of cloud material into ambient material, using the cloud colour field ($C_{\rm c}$). Conversely, we are interested in the mixing of incident `shock' material (SN ejecta) into the cloud. We therefore define the colour-based injection efficiency $\eta^{\rm c}$ as the total mass of SN ejecta in cells containing at least 10 per cent cloud material (i.e. $C_{\rm c} \ge 0.1$), normalized by the initial ejecta mass and the incident ejecta fraction ($\eta_{\rm geom}$). If all of the ejecta incident on the cloud cross-section are `injected' into the cloud, $\eta = 1$.

In the context of Solar system enrichment, we are most interested in enriching the densest (potentially star-forming) regions of the target cloud. Both \citet{2012ApJ...756L...9B} and \citet{2010ApJ...711..597O} consider ejecta to be `injected' above an absolute density threshold. We therefore calculate an alternate injection efficiency, $\eta^{\rm d}$, defined as the total mass of SN ejecta in cells with density greater than the ICM density (i.e. $n > n_{\rm c}$), also normalized by the incident mass fraction ($\eta_{\rm geom}$). This measure only probes the dense clumps; thus if $\eta^{\rm d} \ll \eta^{\rm c}$, most of the ejecta are in diffuse cloud material.

For both measures, we use $\eta_{\rm g}$ for the gas-phase injection and $\eta_{\rm d}$ for general dust grain injection (note that $\eta_{\rm d}$ and $\eta^{\rm d}$ are different quantities). We further determine the dust injection for each radius group, using $\eta_{10}$, $\eta_{1}$, $\eta_{0.1}$, and $\eta_{0.01}$ for the $a =$ 10, 1, 0.1, and 0.01 $\micron$ dust grains, respectively. Further details concerning the injection efficiencies are given in Appendix \ref{a:injection}.

The unknown quantities discussed in Section \ref{ss:dustproduction} (e.g. SLR yield, dust production, dust destruction) can then be included when estimating final SLR abundances. Note that we do not account for radioactive decay during transit. The half-life of $^{26}$Al is $t_{1/2} \approx 0.7$~Myr. For our fiducial SNR, the shock impacts the cloud after roughly 0.03 Myr; therefore only $\sim 3$ per cent of the total ejected $^{26}$Al will have decayed by that time. Over the full duration of our simulation (0.3~Myr), $\sim 25$ per cent of the $^{26}$Al will have decayed. The short half-life of $^{26}$Al underscores the need for both rapid transport and incorporation into the molecular cloud.

\section{Results}\label{s:results}
\subsection{Dynamical evolution}\label{ss:dynamics}

\begin{figure*}
  \includegraphics[width=\textwidth]{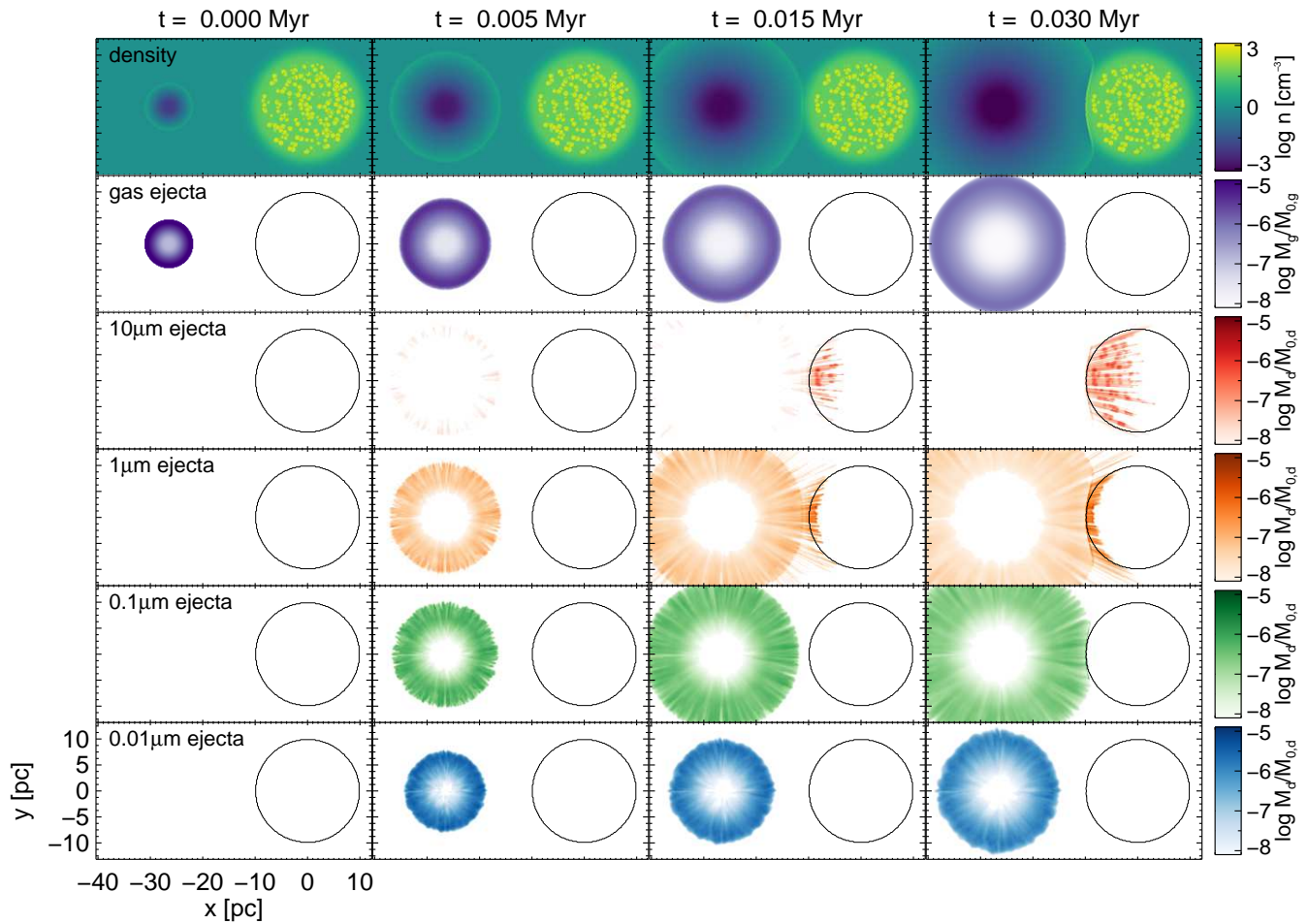} 
  \caption{Time evolution of our fiducial simulation (run F) at early stages (the first 0.03 Myr). Each image is a mid-plane slice at $z=0$. The top row shows the total number density $n$ in ${\rm cm}^{-3}$. All other rows show the mass fraction of each tracer on a per cell basis relative to the initial tracer mass. The second row is the gas-phase ejecta, traced by the colour field $C_{\rm s}$. The remaining rows are the sputtered mass of dust grains from each radius group. The black contour traces the cloud boundary, defined where the cloud colour field $C_{\rm c} \ge 0.1$. The particles located within the central midplane slice ($-\delta_z/2 < z < +\delta_z/2$) are overlaid in grey according to their initial radius group. The smallest grains ($a = 0.01~\micron$) remain well-coupled to the inner ejecta by the drag force and sputter almost completely before impacting the cloud. The $0.1~\micron$ grains outpace the inner ejecta but stall in the forward shock. The large ($a \gtrsim 1.0~\micron$) dust grains decouple and outpace the shock front due to their larger inertia, reaching the cloud and depositing SLRs before the shock impacts the surface. The sputtering of individual particles is visible in the form of radial contrails from the SN centre.}
  \label{f:allsnap1}
\end{figure*}

\begin{figure}
  \includegraphics[width=\columnwidth]{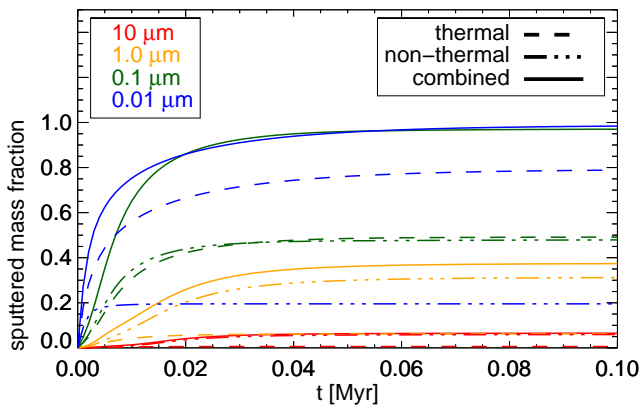} 
  \caption{Fraction of total particle mass eroded by thermal sputtering (dashed), non-thermal sputtering (dash-dot), and the combination of both (solid) during the first 0.1 Myr of the fiducial simulation (run F). colours indicate the initial radius group (red: $10~\micron$; orange: $1~\micron$; green: $0.1~\micron$; blue: $0.01~\micron$). The $0.01~\micron$ grains are rapidly and significantly eroded, predominately by thermal sputtering in the hot SNR. Over 20 per cent of the total mass is lost in the first kyr, and nearly 100 per cent in the first 10 kyr. The $0.1~\micron$ grains also experience rapid destruction but with almost equal contributions from both thermal and non-thermal sputtering, and nearly all are destroyed. The larger grains fare better, with roughly 40 and 10 per cent destruction rates for the $1$ and $10~\micron$ groups respectively. In both instances, the destruction is dominated by non-thermal sputtering as the grains pass through the shock front and into the cold, dense cloud.}
  \label{f:bcsputterplot}
\end{figure}

We follow the evolution of the SN remnant and its interaction with the pre-solar molecular cloud for 0.3 Myr. Fig. \ref{f:allsnap1} shows the first 0.03 Myr of time evolution of the fiducial simulation (run F). As the ST solution is initialized with both kinetic and thermal energy, the pressure discontinuity at the edge of the SNR launches a shock wave (forward shock) into the ambient medium. Because the gas-phase ejecta are traced with a passive colour field ($C_{\rm s}$), they instead follow the contact wave, which lags behind the forward shock. The dust grains begin with the ejecta velocity and therefore initially travel with the expanding gas, experiencing no drag or non-thermal sputtering. However, the high temperatures in the SNR cause significant thermal sputtering. Fig. \ref{f:bcsputterplot} shows the ratio of sputtered mass to total mass for each grain radius group over time. At early times, thermal sputtering dominates and erodes nearly 80 per cent of the smallest ($a=0.01~\micron$) grains.

As the remnant expands into the ambient medium, the forward shock accumulates more material, eventually slowing and cooling into a dense shell. The smallest grains ($a = 0.01~\micron$) remain well-coupled to the inner gas ejecta. Slightly larger ($a = 0.1~\micron$) grains outpace the inner ejecta but stall in the dense forward shock. The relative velocity difference then generates non-thermal sputtering, which contributes almost equally to the destruction of the $0.1~\micron$ grains (compare the dashed and dash--dotted green lines in Fig. \ref{f:bcsputterplot}). Both of the smaller grain groups are almost completely stopped and destroyed by sputtering within the remnant. In contrast, the larger grains ($a \ge 1.0~\micron$) remain largely intact and dynamically decouple from the ejecta due to their higher inertia. The large grains also pass through the forward shock and ballistically impact the cloud before the shock arrives. Once in the cloud, the grains rapidly slow and kinetically sputter due to the increased densities and high relative velocities.

The behaviour of the dust grains in the SN remnant agrees well with the results of \citet{2016A&A...587A.157B}. The authors performed 1D simulations of the growth and erosion of dust in SNRs including multiple grain compositions, plasma drag, and detailed sputtering. Despite using simplified dust physics, we obtain very similar results to the evolution of Mg$_2$SiO$_4$ presented in fig.~3 of \citet{2016A&A...587A.157B}: (1) small grains ($a = 0.01~\micron$) are highly eroded in the remnant and remain within the ejecta region; (2) slightly larger ($a = 0.1~\micron$) grains pass through the ejecta but remain within the forward shock; and (3) the larger ($a = 1.0~\micron$) grains are eroded very little and eventually move beyond the forward shock.

\begin{figure*}
  \includegraphics[width=\textwidth]{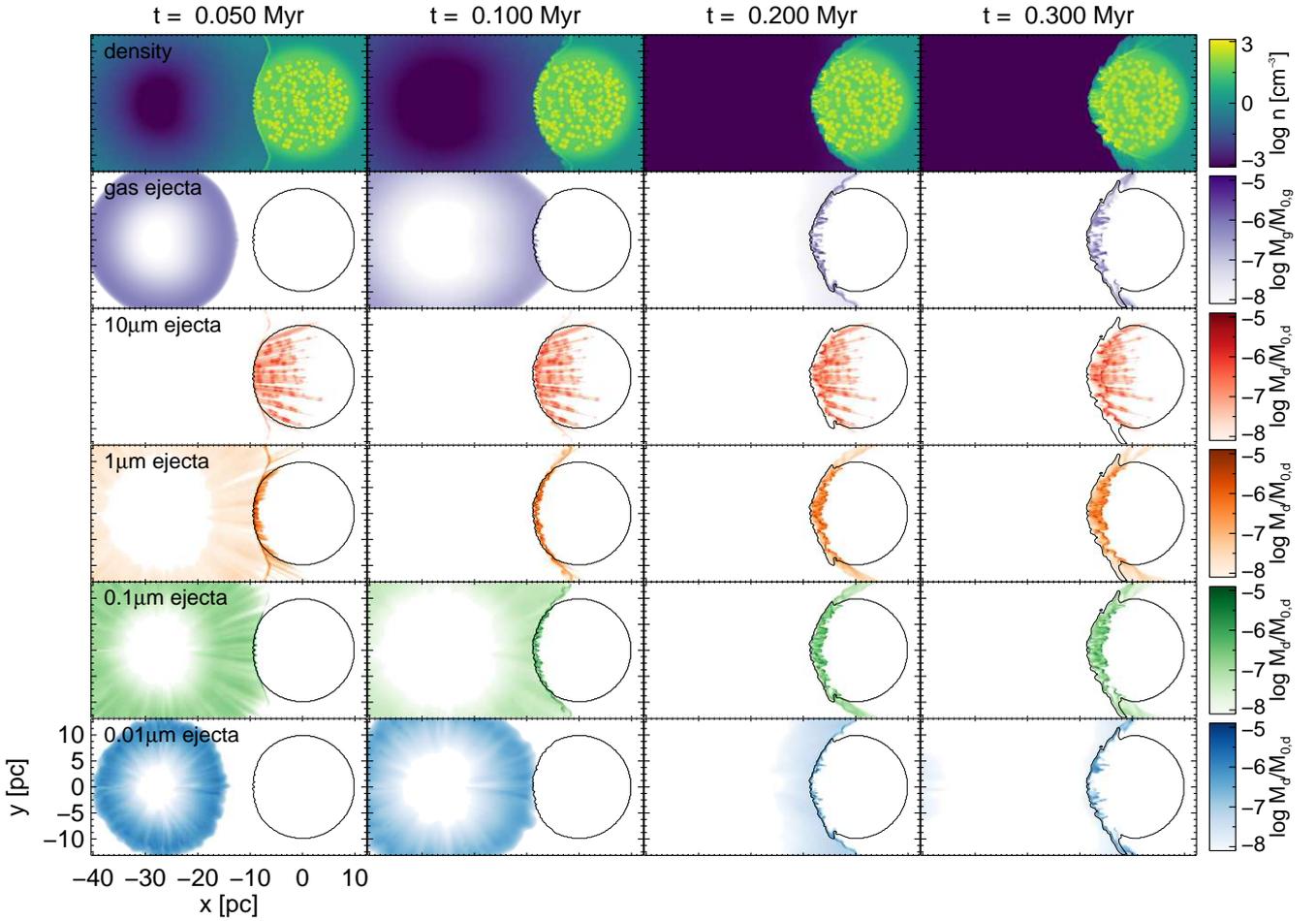} 
  \caption{Same as Fig. \ref{f:allsnap1}, but at later stages. The forward shock impacts the cloud within $\sim$ 0.03~Myr, but the inner ejecta does not arrive until $t \sim 0.06$~Myr. The clumpy substructure of the cloud creates channels for the impinging gas to penetrate and mix. At later times, Rayleigh--Taylor instabilities lead to injection of gaseous SLRs through the cloud surface. After 0.3~Myr, nearly all the grains within the cloud have either been stopped or sputtered. Nearly half of the dust grains incident on the cloud are captured, and the largest grains penetrate furthest.}
  \label{f:allsnap2}
\end{figure*}

Fig. \ref{f:allsnap2} shows the evolution of the simulation after forward-shock impact. As noted in Section \ref{ss:thermal}, the SNR is only just starting to cool when it impacts the molecular cloud surface. The expansion velocity of the shell is still supersonic ($\sim 350$~km~s$^{-1}$) at impact. The hot, diffuse gas encounters a cold, dense wall and deflects around the edges, ablating material. A slower shock is transmitted into the cloud, and the clumpy substructure provides channels and gaps for the gas to enter the cloud. Both the clumpy substructure and the efficient radiative cooling prevent the formation of a stand-off shock, which is usually observed in the adiabatic shock--cloud interaction \citep{2006ApJS..164..477N} and could drastically limit the SLR injection (see Section \ref{ss:effcool}). At late times, the Rayleigh--Taylor instability begins to manifest at the cloud surface, driving fingers into the cloud that will eventually mix and inject SLRs in the gas phase.

\begin{figure*}
  \includegraphics[width=\textwidth]{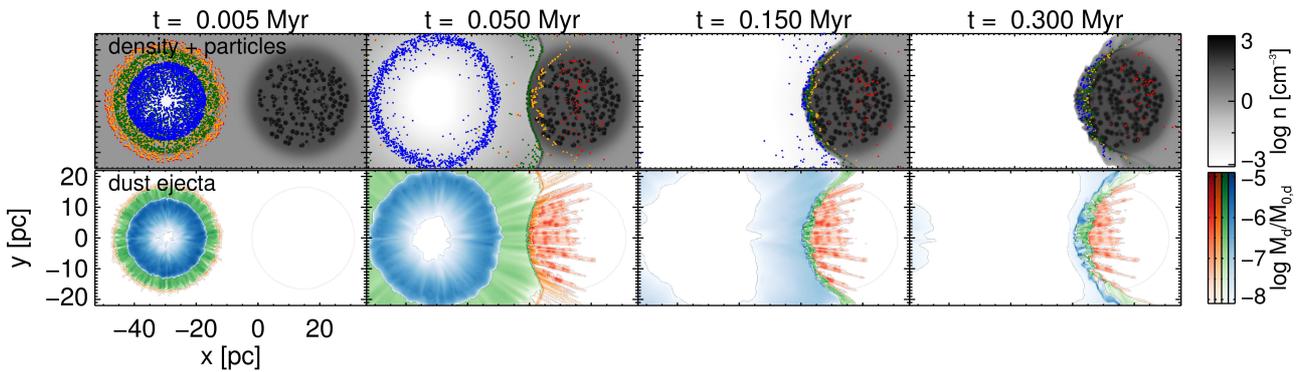} 
  \caption{Time evolution of the fiducial simulation (run F) illustrating the spatial stratification of the dust grains (top row) and sputtered SLRs (bottom row) due to initial grain radius distribution. As in Figs \ref{f:allsnap1} and \ref{f:allsnap2}, the images are mid-plane slices at $z=0$. In the top row, particles located from $-\delta/2 \le z \le +\delta/2$ are overlaid on a desaturated map of number density. Each dust grain group is colour-coded by initial radius (red: $10~\micron$; orange: $1~\micron$; green: $0.1~\micron$; blue: $0.01~\micron$). The same colour scheme holds in the bottom row, now showing the sputtered SLR mass fraction, relative to the initial tracer mass. The grey contour defines the cloud edge. As the simulation proceeds, the dust grains separate spatially based on initial radius, with the larger grains travelling further into the cloud. This stratification could help explain anomalies in observed Solar system abundances, such as the low $^{60}$Fe/$^{26}$Al ratio.}
  \label{f:allparsnap}
\end{figure*}

In contrast to the hydrodynamical (gas-phase) mixing, the large dust grains rapidly inject SLRs throughout the cloud. Fig. \ref{f:allparsnap} shows the evolution of the dust grains, as well as a combined view of the sputtered mass from each initial radius group. The largest ($a = 10~\micron$) grains penetrate furthest, sputtering most of their mass in the leading edge of the cloud. The smaller grains have been largely stopped and sputtered before entering the cloud. Still, the SLR contents of the $0.1~\micron$ grains have outpaced the inner ejecta and mix into the dense gas $\sim 0.05$ Myr earlier. Nearly all grains incident on the cloud are sputtered and stopped within the cloud, i.e. only grains at grazing angles can re-emerge from the cloud interior. 

\subsection{Injection of SLRs}\label{ss:injecteff}

\begin{figure}
  \includegraphics[width=\columnwidth]{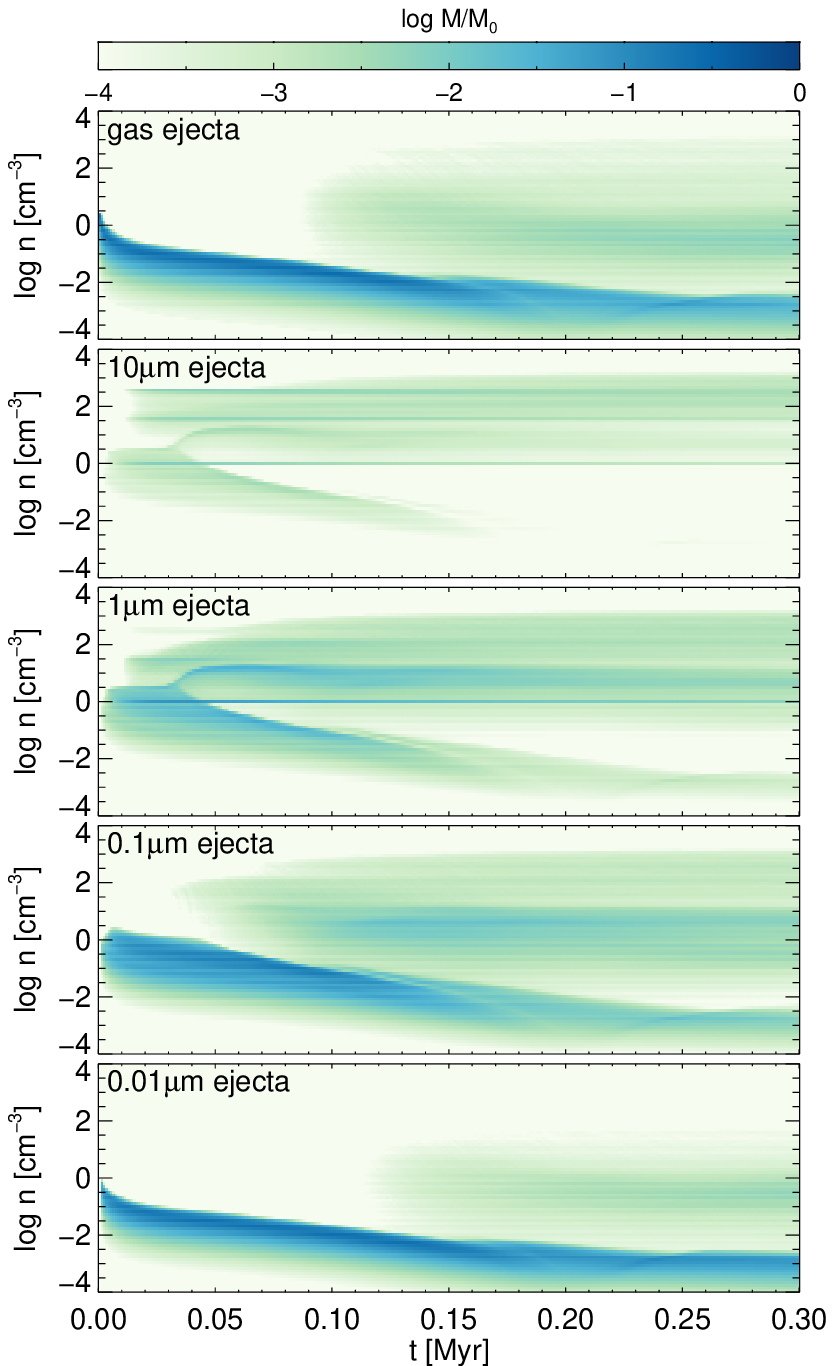} 
  \caption{Tracer mass fraction binned logarithmically by density across time for the fiducial simulation (run F). The top panel shows the SNR gas tracer ($\rho C_{\rm s}$). The rest of the panels show the mass deposited into gas phase by dust grain sputtering ($\rho_{\rm d}$) for each initial grain size (10, 1, 0.1, and 0.01~$\micron$). While hydrodynamical mixing is largely restricted to later times and low cloud densities (top panel), the large ($a \ge 1~\micron$) dust grains enrich higher densities at earlier times. Only the smallest grains do not reach higher densities. The prominent horizontal line at $n = 1~{\rm cm}^{-3}$ corresponds to the ambient medium, while lower densities are located in the diffuse SNR.}
  \label{f:bcinjectfrac}
\end{figure}

\begin{figure}
  \includegraphics[width=\columnwidth]{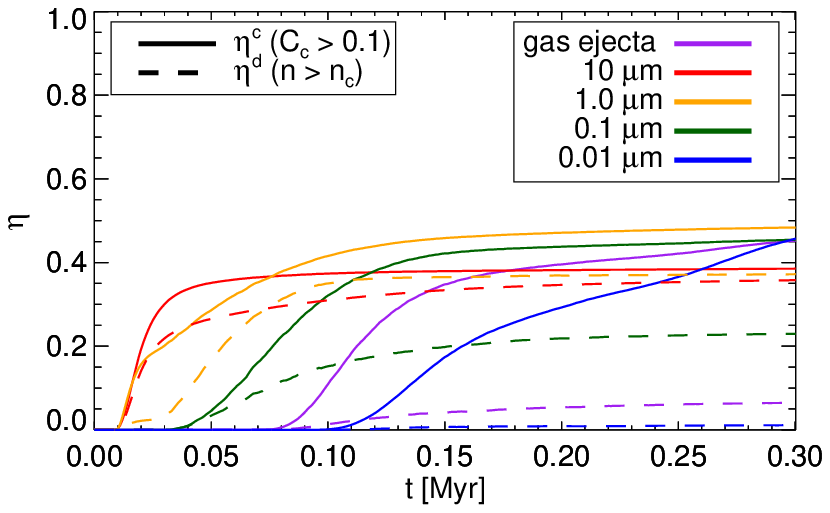} 
  \caption{Injection efficiency $\eta$ as a function of time for the ejecta in our fiducial simulation (run F). Injection is measured using the cloud tracer (solid) and density threshold (dashed). Each tracer is colour-coded as in Fig. \ref{f:allsnap1} (purple: gas; red: $10~\micron$; orange: $1~\micron$; green: $0.1~\micron$; blue: $0.01~\micron$). The largest grains ($a \ge 1~\micron$) arrive within the first 0.01 Myr and rapidly deposit a substantial fraction ($\gtrsim 20$ per cent) of their SLR mass within 0.1 Myr. The intermediate grains ($a = 0.1~\micron$) are sputtered and stopped in the forward shock and arrive slightly ahead of the gas. The smallest grains ($a = 0.01~\micron$) sputter significantly before entering the cloud, yet injection of SLRs from these grains continues as gas at the leading edge of the cloud is subsequently mixed by Rayleigh-Taylor instabilities. While the colour-based injection is approximately the same ($\eta^{\rm c} \gtrsim 0.4$) for all ejecta types, the density-based injection ($\eta^{\rm d}$) decreases with grain radius, indicating most of the smaller grain deposition is in diffuse intercloud gas.}
  \label{f:bcmixfracplot}
\end{figure}

We are interested in the enrichment of the densest (eventually star-forming) gas. Therefore, we analyse the SLR deposition as a function of density (Fig. \ref{f:bcinjectfrac}). Comparing the dust ejecta to the gas ejecta, the gas ejecta are mostly distributed in the diffuse SNR and background ISM. In contrast, the $10~\micron$ grains deposit a significant fraction of mass into the densest gas, and smaller particles deposit smaller fractions in the dense gas. This effect is further quantified in Fig. \ref{f:bcmixfracplot}, which compares the injection efficiency $\eta$ of both dust and gas as a function of time. At late times, the colour-based injection efficiency is roughly equivalent for all grain sizes ($\eta^{\rm c} \sim 0.5$), indicating that nearly half the incident material has been mixed into the cloud. However, the density-based injection $\eta^{\rm d}$ decreases with decreasing grain size, to the point that the smallest grains and gas deposit only negligible amounts of ejecta in the densest regions. This agrees qualitatively with \citet{2012ApJ...756L...9B}, who found only a small fraction of incident gas-phase material is injected into a dense pre-stellar core ($\eta^{\rm d}_{\rm g} \approx 0.03$). This indicates that only the large grains are able to enrich the densest gas ($n > n_{\rm c}$). Table \ref{tab:results} provides a summary of final injection efficiencies from all simulations performed.

Because the SLRs decay, the enrichment needs to occur rapidly. As seen in both Figs \ref{f:bcinjectfrac} and \ref{f:bcmixfracplot}, the particles are able to deposit SLRs in the cloud $\sim 0.1$ Myr before gas phase mixing occurs. For all dust grain sizes, the injection of SLRs occurs rapidly, reaching peak values in less than 0.1 Myr. This is contrasted with the gas, which slowly mixes and is still increasing its injection amount when the simulation ends. The gas injection efficiency only becomes comparable to the dust injection efficiencies after 0.2 Myr.

\subsection{Resolution convergence}\label{ss:resolution}
In inviscid hydrodynamics, the degree of mixing is controlled by numerical viscosity. As the numerical diffusion decreases with increasing resolution, adequate resolution is necessary to properly capture the dynamics. In the two-dimensional shock-cloud interaction, previous work has found that about 100 cells per cloud radius ($N_R \gtrsim 100$) are necessary for convergence of global quantities \citep{1994ApJ...420..213K,2006ApJS..164..477N}. This requirement may be reduced in three-dimensional simulations to $N_R \gtrsim 32-64$ \citep{2016MNRAS.457.4470P}, and our fiducial simulation falls within this range ($N_R \approx 50$). However, because the instabilities tend to grow fastest on the smallest scales, the details of the small-scale mixing could be dominated by resolution effects. \citet{2008ApJ...680..336S} found that all quantities except the cloud mixing fraction show convergence, and our definition of the injection efficiency is similar to their mixing fraction. 

\begin{figure}
  \includegraphics[width=\columnwidth]{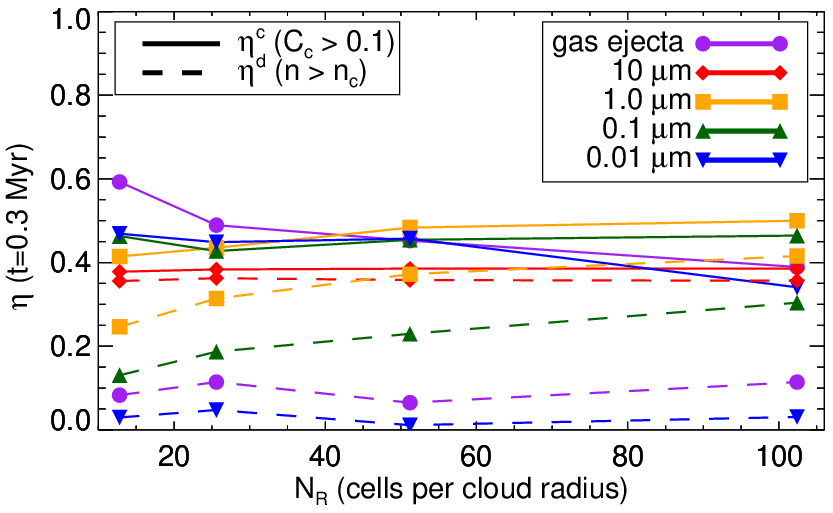} 
  \caption{Injection efficiencies $\eta$ as a function of simulation resolution, represented by the cells per cloud radius $N_R$. $\eta$ is evaluated at simulation termination ($t=0.3$ Myr) using the cloud tracer (solid) and density threshold (dashed). Each tracer is colour-coded as in Fig. \ref{f:allsnap1} (purple: gas; red: $10~\micron$; orange: $1~\micron$; green: $0.1~\micron$; blue: $0.01~\micron$). The injection efficiency of the largest grains increases slightly for both measures, as the density peaks within the clumps are better resolved and capture more material. Injection decreases for the smallest grains and the gas due to decreased numerical diffusion at the cloud surface.}
  \label{f:bcresplot}
\end{figure}

Fig. \ref{f:bcresplot} compares the fiducial result to simulations performed at both lower and higher resolution (runs R1--R4), up to 100 cells per radius ($1024\times512\times512$ grid points). The injection efficiency of the larger grains ($a \ge 0.1~\micron$) increases only slightly with increasing resolution. In contrast, the injection of smallest grains ($a = 0.01~\micron$) and the gas ejecta decreases as the resolution increases. The larger injection efficiencies at lower resolution may be attributable to increased numerical diffusion, leading to increased mixing at the cloud interface. Overall, the trend is sufficiently flat to conclude our three-dimensional simulations are well-resolved at $N_R = 50$, in agreement with \citet{2016MNRAS.457.4470P}.

In the previous resolution test, we kept the number of particles fixed at $N_{\rm p} = 10^5$. We do not expect the particles to be strongly affected by simulation resolution. However, the number of particles used may alter the injection. As the particles are placed randomly within the SNR, a sufficient number of particles are required to eliminate any gaps when the shock wave encounters the cloud surface. We repeat our fiducial simulation varying the number of particles from $N_{\rm p} = 10^4$ (run N1) to $N_{\rm p} = 10^6$ (run N2). We find no significant variation in injection efficiency with particle number (see Table \ref{tab:results}). 

\begin{figure*}
  \includegraphics[width=\textwidth]{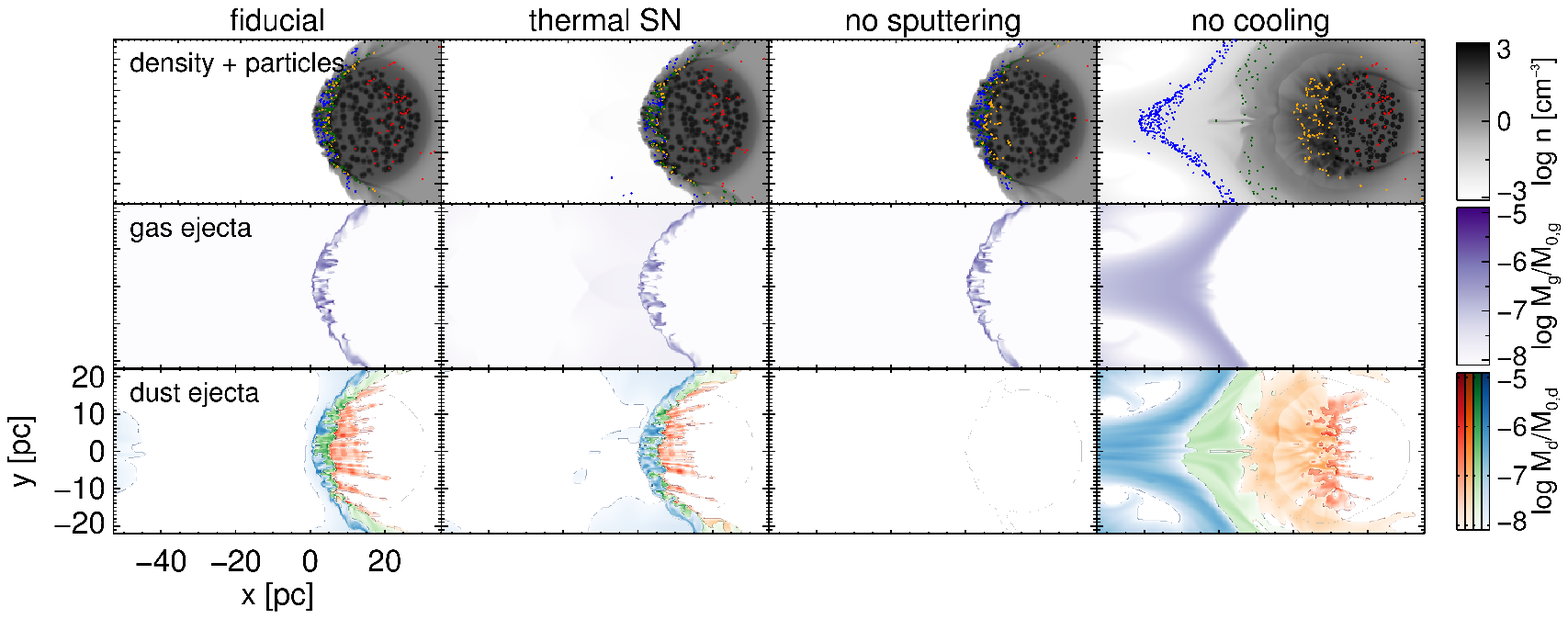} 
  \caption{Similar to Fig. \ref{f:allparsnap}, but comparing different simulations at $t = 0.3$~Myr. From left to right, the columns show (a) the fiducial simulation (run F); (b) the SNR initialized as a thermal pulse rather than a ST solution (run FT); (c) no sputtering of dust grains (run NS); and (d) no thermal physics, i.e. purely adiabatic with no heating or cooling (run NC). The middle row shows the gas ejecta tracer field. Comparing the fiducial to the thermal pulse, the initialization of the SNR does not appear to drastically alter the evolution or injection. Without sputtering, there are no SLRs released in gas phase; hence the bottom panel is blank. The $10~\micron$ dust grains are not stopped in the cloud by drag and re-emerge. Without cooling, the impact of the SNR creates a bow shock that deflects incoming gaseous ejecta. The large, intact grains still decouple and penetrate the cloud surface, injecting SLRs.}
  \label{f:mdlparsnap}
\end{figure*}

\subsection{Effect of supernova remnant model}\label{ss:effsnr}
We have also performed a simulation using the standard thermal pulse to initialize the SNR rather than an exact ST solution. In this model (run FT), we inject $E_{\rm SN} = 10^{51}$~erg of thermal energy and $M_{\rm ej} = 10$~M$\odot$ uniformly into a spherical volume of radius 20 cells. With sufficient resolution, this approach has been shown to evolve approximately into the ST solution after only 2~kyr \citep{2015ApJ...802...99K} and is therefore often used for its simplicity \citep{2013ApJ...769L...8V}. Because the SNR has no initial kinetic energy, injecting the particles at the start of the simulation would generate non-physical drag. We therefore let the thermal pulse evolve for 3~kyr before injecting the particles, which are then placed in the forward shock with the local gas velocity.

The final state of the simulation is displayed in the second column of Fig. \ref{f:mdlparsnap}. Overall, the result obtained using the thermal pulse is almost indistinguishable from the ST model -- the shock thickness, velocity, and arrival time are approximately the same, and the injection efficiencies at simulation termination are nearly identical (see Table \ref{tab:results}). There is a small difference in the dust grains due to the initialization; because we wait 3~kyr to insert the dust in the thermal pulse model, the grains experience less thermal sputtering and arrive later.

\subsection{Effect of sputtering}\label{ss:effsput}
We compare the fiducial results to a simulation run without sputtering (run NS). The third column of Fig. \ref{f:mdlparsnap} shows the result at simulation termination. The hydrodynamics and the gas tracer field are not affected by the lack of sputtering, since there is no feedback from the particles to the gas. The drag force depends on the dust grain radius, and therefore the dust dynamics are altered by the lack of sputtering. The largest grains pass almost entirely through the target cloud. The rest of the grains also travel further into the cloud but are eventually stopped by the drag force. Since there is no sputtering, no SLRs are released into the gas phase and the injection is measured solely by the stopped grain criterion (see Appendix \ref{a:injection}). 

\subsection{Effect of radiative cooling}\label{ss:effcool}
We compare the fiducial results to a simulation run without radiative heating and cooling (run NC). As seen in the fourth column of Fig. \ref{f:mdlparsnap}, the behaviour of the gas is radically altered. As the shock wave impacts the cloud surface, the purely adiabatic equation of state results in the formation of a stand-off shock at the leading edge of the cloud, diverting impinging material and preventing mixing. The gas-phase ejecta and the smaller grains (which are either coupled to the gas or sputtered) do not mix at all with cloud material, and the injection efficiency is essentially zero (see Table \ref{tab:results}). Cooling lowers the effective adiabatic index of the shock-cloud interaction. As the gas is compressed, the strong radiative losses reduce the shock stand-off distance, allowing mixing of phases and enhancing injection. The larger dust grains are less affected as they are largely intact at impact and still penetrate the cloud surface.

\subsection{Filling factors}\label{ss:ff}
In contrast to previous shock--cloud and SN injection simulations, we include substructure in the target cloud through high-density clumps randomly embedded in an ICM. The fiducial simulation has a cloud volume filling factor of $\phi = 0.5$. We expect the SN shock wave to interact differently as the filling factor is varied. Clumps at the cloud surface provide channels for injection, reducing the need for Rayleigh--Taylor fingers. We examine the effect of varying the filling factor from $\phi = 0.1$ to $0.9$ (runs F1--F9); results are given in Table \ref{tab:results}. Overall, the dust grain injection is largely unaffected by the filling factor. A higher $\phi$ leads to slightly increased injection efficiency, most notably in the $10~\micron$ grains, as the additional clumps capture grains on trajectories near the periphery of the cloud.

\begin{table*}
  \centering
  \caption{Summary of simulations and results.}
  \label{tab:results}
  \begin{tabular}{cccccccccccccc}
    \hline
    Run name & $\phi$ & $N_{R}$ & $N_{\rm p}$ & $\eta^{\rm c}_{\rm g}$ & $\eta^{\rm c}_{10}$ & $\eta^{\rm c}_{1}$ & $\eta^{\rm c}_{0.1}$ & $\eta^{\rm c}_{0.01}$ & $\eta^{\rm d}_{\rm g}$ & $\eta^{\rm d}_{10}$ & $\eta^{\rm d}_{1}$ & $\eta^{\rm d}_{0.1}$ & $\eta^{\rm d}_{0.01}$ \\
    \hline
    F  & 0.5 & 50 & 10$^5$ & 0.45 & 0.39 & 0.48 & 0.45 & 0.46 & 0.065 & 0.36 & 0.37 & 0.23 & 0.011  \\
    R1 & 0.5 & 12 & 10$^5$ & 0.59 & 0.38 & 0.42 & 0.46 & 0.47 & 0.083 & 0.36 & 0.25 & 0.13 & 0.030  \\
    R2 & 0.5 & 25 & 10$^5$ & 0.49 & 0.38 & 0.44 & 0.43 & 0.45 & 0.11  & 0.36 & 0.31 & 0.19 & 0.048  \\
    R4 & 0.5 & 100& 10$^5$ & 0.39 & 0.39 & 0.50 & 0.46 & 0.34 & 0.11  & 0.36 & 0.42 & 0.30 & 0.031  \\
    N1 & 0.5 & 50 & 10$^4$ & 0.44 & 0.38 & 0.47 & 0.44 & 0.44 & 0.066 & 0.35 & 0.37 & 0.23 & 0.012  \\
    N2 & 0.5 & 50 & 10$^6$ & 0.45 & 0.38 & 0.47 & 0.44 & 0.44 & 0.066 & 0.35 & 0.36 & 0.22 & 0.011  \\
    FT & 0.5 & 50 & 10$^5$ & 0.30 & 0.41 & 0.51 & 0.46 & 0.29 & 0.045 & 0.38 & 0.41 & 0.31 & 0.056  \\
    NS & 0.5 & 50 & 10$^5$ & 0.45 & 0.00 & 0.0012 & 0.011 & 0.029 & 0.065 & 0.00 & 0.0012 & 0.0096 & 0.011  \\
    NC & 0.5 & 50 & 10$^5$ & 0.00 & 0.37 & 0.30 & 0.00033 & 0.00 & 0.00 & 0.29 & 0.043 & 0.000 & 0.000 \\
    F1 & 0.1 & 50 & 10$^5$ & 0.46 & 0.34 & 0.49 & 0.45 & 0.47 & 0.052 & 0.23 & 0.36 & 0.22 & 0.0081 \\
    F3 & 0.3 & 50 & 10$^5$ & 0.45 & 0.37 & 0.48 & 0.45 & 0.46 & 0.057 & 0.32 & 0.36 & 0.22 & 0.0092 \\
    F7 & 0.7 & 50 & 10$^5$ & 0.43 & 0.40 & 0.48 & 0.44 & 0.44 & 0.071 & 0.38 & 0.38 & 0.23 & 0.012  \\
    F9 & 0.9 & 50 & 10$^5$ & 0.42 & 0.40 & 0.48 & 0.44 & 0.42 & 0.071 & 0.39 & 0.39 & 0.24 & 0.010  \\
    \hline
  \end{tabular}
\end{table*}

\section{Discussion}\label{s:discussion}

\subsection{The Role of dust grains}
We investigate the role of SN dust grains in enriching a nearby molecular cloud with SLRs. Our results indicate that dust grains formed in SN ejecta can survive transport through the ISM and significantly enrich neighbouring clouds. We find that sufficiently large grains ($a \ge 1~\micron$) decouple from the expanding SN remnant, pass through the shock front and cloud surface, and deposit a significant fraction of their SLR mass into the cloud ($\eta_{\rm d} \gtrsim 0.4$). In particular, large dust grains enrich the dense gas rapidly, preventing significant SLR decay. Smaller grains ($a \le 0.1~\micron$) sputter and stall in the SNR and contribute SLRs predominately through gas-phase mixing. The gas-phase ejecta mix slowly through hydrodynamic and thermal instabilities at the cloud surface.

Our results agree with those of \citet{2010ApJ...711..597O} despite using very different targets (molecular cloud versus proto-stellar disc) and SN distances (18~pc versus 0.1--2 pc). The authors found that a considerable fraction ($\eta \gtrsim 0.8$) of grains larger than $1~\micron$ are injected into the target, which compares favorably with our estimates ($\eta_{\rm d} \gtrsim 0.4$). Similarly, the smallest grains ($a = 0.01~\micron$) are slowed and completely destroyed. We also find approximate agreement with our estimate for gaseous injection; \citet{2010ApJ...711..597O} estimated $\eta_{\rm g} \lesssim 0.01$, while we find $\eta^{\rm d}_{\rm g} \approx 0.1$.

\subsection{$^{60}$Fe/$^{26}$Al ratio}\label{ss:feal}
We observe that the dust drag and sputtering naturally lead to a spatial stratification between grains of different sizes, illustrated in Fig. \ref{f:allparsnap}. One of the leading arguments against a SN enrichment source is that the SLR abundances in our Solar system do not precisely match predicted SN yields. In particular, some estimates of the ratio of $^{60}$Fe/$^{26}$Al in the ESS are orders of magnitude lower than expected in SNe \citep{2012E&PSL.359..248T}, casting doubt on a SN origin for ${26}$Al \citep{2015A&A...582A..26}. However, if the primary carriers of $^{60}$Fe and $^{26}$Al condense into grains of different characteristic radii, these isotopes may not end up in the same dense gas reservoirs. In addition, the ejecta of SNRs are not spatially homogeneous \citep{2014Natur.506..339G}. Both observational \citep{2010ApJ...725.2038D} and simulation \citep{2015A&A...577A..48W} results indicate that iron-group elements may be preferentially ejected in a particular direction. If the pre-solar cloud was not in this narrow window, it would receive far less $^{60}$Fe than predicted, and a SN may still be the injection source.

\subsection{Other considerations}
We do not consider the evolution of the SN progenitor prior to explosion. The progenitor's stellar wind and ionizing radiation will shape the circumstellar environment, resulting in a stratified medium ($\rho \propto r^2$) rather than a uniform medium. This density gradient will affect the transit of the shock wave and grains through the intervening gas. Furthermore, the stellar wind will contain dust grains that may also be enriched with certain SLRs, such as $^{26}$Al, produced during main sequence and post-main sequence evolution \citep{2006ApJ...647..483L,2005A&A...429..613P}. These enriched dust grains will be swept up by the passage of the subsequent SNR and may further enhance SLR enrichment \citep{2012A&A...545A...4G}.

We consider only one set of parameters for the SN (explosion energy $E_{\rm SN}=10^{51}$~erg and ejected mass $M_{\rm ej} = 10~{\rm M}\odot$) at a single distance ($d = 18$~pc). The SN parameters are somewhat constrained and only slightly affect the initial condition. The SN distance is limited by the estimated SLR yield of SNe, the geometric dilution of ejecta, and the radioactive decay of SLRs. As noted in Section \ref{sss:snr}, our chosen separation is at the upper limit of the `radioactivity distance' for $^{26}$Al enrichment \citep{2006ApJ...652.1755L}. Reducing the distance from the SN to the pre-solar cloud may increase injection due to decreased geometric dilution, increased shock speed at impact, decreased time for radioactive decay of SLRs, and decreased sputtering. Therefore our estimates may be considered a lower limit in this regard.

We also only consider a single SN. However, most massive stars form in clustered environments, e.g. OB associations \citep{2003ARA&A..41...57L}, and in multiple systems \citep{2007ARA&A..45..481Z}. Indeed, it is likely that multiple SNe over one or more generations contributed SLRs to the pre-solar cloud \citep{2013ApJ...769L...8V,2014E&PSL.392...16Y}.

Cloud morphology may also play a considerable role in gas injection. We introduce static, clumpy substructure in the target cloud. The substructure prevents a symmetric stand-off shock from forming after impact and provides diffuse channels for injection through the dense filaments. The break-up of the shock also generates turbulence and mixing. We have neglected dynamical perturbations (velocity substructure); however, molecular clouds are probably turbulent \citep{2004ARA&A..42..211E}, and introducing turbulence could further enhance the mixing at the cloud surface and increase injection of the smaller grains and gas.

We do not include gravity in our simulations. The potential effect of gravity can be estimated by comparing the local free-fall time $t_{\rm ff} = [3 \pi / (32 G \rho)]^{1/2}$ to the simulation time. For the dense clumps with $n_{\rm cl} \approx 400~{\rm cm}^{-3}$, $t_{\rm ff} \approx 2$~Myr -- much longer than the time-scales considered here (0.3~Myr). However, we note that compression by the SN shock wave, as well as fragmentation due to thermal instability, will create higher densities and may trigger collapse. Due to the global nature of our simulation, we are limited to measuring injection efficiencies at large scales within the cloud. Following the enrichment and mixing down to individual pre-stellar cores (sub-parsec scale) will require gravity and additional resolution (possibly through mesh refinement). While the densest gas is harder to penetrate, collapsing cores could receive SLRs by accreting enriched diffuse gas during collapse \citep{2016ApJ...826...22K}.

In our dust drag law, we consider only neutral grains and ignore the Coulomb drag force [second term in equation 4 of \citet{1979ApJ...231...77D}]. However, dust grains will be charged by collisions with ions \citep{1979ApJ...231...77D}. The Coulomb term will become large when the relative velocity approaches the sound speed and may significantly affect the grain dynamics at low relative velocities. Within the SNR, the dust-to-gas relative velocity is low and the gas temperature is high; hence the Coulomb (plasma) drag may be several times larger than the collisional drag \citep[fig.~2, ][]{2016A&A...587A.157B}. Reducing the dust grain velocities may reduce the injection, and therefore our estimates of enrichment may be upper limits. Charged grains will also interact with any magnetic fields present in the gas, which we neglect. Within the SN ejecta, the dust grains may be largely unaffected by magnetic effects, as there is observational and numerical evidence that the field is radially aligned \citep{2009MNRAS.394.1307D,2013AJ....145..104R,2013ApJ...772L..20I}. However, the field orientation may shift at the SN shock front; as noted by \citet{1997ApJ...489..346F}, gas--grain de-coupling may be suppressed or even prevented by fields in the shock front, which could drastically reduce the enrichment. Magnetic effects could also alter the grain dynamics within the target cloud. The average magnetic field increases with column density in dense molecular gas \citep{2012ARA&A..50...29C}; hence the effect on grains also increases near star-forming clumps. Future work on the subject should consider the combined effects of grain charging, Coulomb drag, and magnetic fields.

\section{Conclusions}\label{s:conclusions}
A nearby SN remains a possible candidate as the source of SLRs in the early Solar system. The main challenge in this `direct injection' scenario is overcoming the impedance mismatch between the hot, diffuse SNR gas and the cold, dense pre-solar gas, as demonstrated amply in the literature \citep{2012ApJ...756L...9B,2012ApJ...745...22G,2012ApJ...756..102P}. We explore whether dust grains formed from the SN ejecta and carrying SLRs can overcome the mixing barrier and enrich dense (potentially star-forming) gas. Using hydrodynamical simulations, we model the interaction of a SNR carrying dust grains with the pre-solar molecular cloud. We follow dust grains of varying initial radius ($a = 0.01$--$10~\micron$) subject to drag forces and sputtering. We find the following points:
\begin{enumerate}
\item Sufficiently large dust grains ($a \ge 1~\micron$) entrained in the SN ejecta will decouple from the shock front and survive entry into the molecular cloud. They will then be either completely stopped or sputtered, enriching the dense gas with SLRs within 0.1 Myr of the SN explosion.
\item Smaller dust grains ($a \le 0.1~\micron$) formed in the SN ejecta will be either stopped or sputtered before impacting the molecular cloud. The sputtered SLRs will contribute to the enrichment through subsequent gas-phase mixing.
\item Gas-phase SN ejecta will enrich the leading edge of molecular cloud only after instabilities develop at the cloud surface. The degree of mixing depends strongly on the inclusion of radiative cooling.
\end{enumerate}
While it is still unknown what fraction of dust grains survive passage by the reverse shock and emerge from the SNR, we show that any surviving dust will contribute favorably to the typical SN enrichment scenario. Indeed, if a significant amount of large ($a \gtrsim 1~\micron$) grains survive, dust may be the dominant source of SLR enrichment in nearby molecular clouds. Most notably, the dust grain enrichment occurs rapidly, in contrast with the typical gas-phase mixing which relies on the growth of hydrodynamical instabilities at the cloud surface. A shorter time delay between production and injection of the SLRs prevents substantial radioactive decay. Finally, if the various SLRs condense into different-sized dust grains, drag and sputtering will lead to a spatial stratification of SLRs within the pre-solar cloud. This could explain the large discrepancy in the $^{60}$Fe/$^{26}$Al mass ratio between SN predictions and meteoritic measurements.  We conclude that dust grains can be a viable mechanism for the transport of SLRs into the pre-solar cloud.

\section*{Acknowledgements}
We thank the referee, Marco Bocchio, for an insightful and helpful report. MDG thanks Jim Stone, Eve Ostriker, Bruce Draine, and Jonathan Tan for helpful discussions. Computations were performed on the Kure Cluster at UNC-Chapel Hill. MDG and FH gratefully acknowledge support by NC Space Grant and NSF Grant AST-1109085. IL acknowledges support from NSF Award ACI-1156614, which supported the UNC Chapel Hill Computational Astronomy and Physics Research Experience for Undergraduates programme.

%%%%%%%%%%%%%%%%%%%%%%%%%%%%%%%%%%%%%%%%%%%%%%%%%%

%%%%%%%%%%%%%%%%%%%% REFERENCES %%%%%%%%%%%%%%%%%%

\bibliographystyle{mnras}

%%%%%%%%%%%%%%%%%%%%%%%%%%%%%%%%%%%%%%%%%%%%%%%%%%

%%%%%%%%%%%%%%%%% APPENDICES %%%%%%%%%%%%%%%%%%%%%

\appendix

\section{Modifications to \textsc{Athena}}\label{a:mods}
We have added passive particles to the VL integrator \citep{2009NewA...14..139S} in \textsc{Athena} \citep{2008ApJS..178..137S}. The particle update is performed using the predictor values. Comparisons with the CTU integrator, which includes particles by default, show nearly absolute agreement. To include sputtering, we have modified the particle implementation to track individual particle radii. 

\section{Dual energy formulation}\label{a:dualenergy}
In regions of high kinetic energy, the calculation of internal energy via subtraction from the total energy can lead to negative pressures, specifically during reconstruction. We have therefore implemented a procedure nearly identical to that described in section 4.1.1 of \citet{2014ApJS..211...19B}\footnote{We write the internal energy density $e$, which is equivalent to $\rho e$ in the notation of \citet{2014ApJS..211...19B}}. We simultaneously solve the internal energy equation (equation \ref{eq:eint}) for every cell in our domain. In cells where the internal energy is a small fraction of the total energy, $(E - \rho|{\bf v}|^2/2)/E \le 10^{-3}$, we revert to using $e$, as in equation (44) of \citet{2014ApJS..211...19B}. This check is performed any time the internal energy (or pressure or temperature) are required, such as calculating the pressure at cell interfaces as inputs to the Riemann solver. We prefer the dual energy formulation over a pressure or temperature floor in our models; while reverting the pressure to a small number ($\sim 10^{-20}$) may not affect the dynamics in most situations, the cooling depends very sensitively on the temperature.

The internal energy equation (equation \ref{eq:eint}) is not conservative. The left-hand side can be treated as an advection equation for $e/\rho$. We therefore use the density flux returned from the Riemann solver to advect the internal energy, treating $e$ as a passive colour field. The source term is calculated and applied at cell centres using a monotonic central difference to evaluate the gradients of the velocity in each direction. In contrast to \citet{2014ApJS..211...19B}, we use the updated pressure [calculated from $P = e (\gamma-1)$] when applying the source term at the full time-step update in the VL integrator. The non-conservative formulation can lead to large discrepancies from the correct internal energy if the equation is allowed to evolve on its own. Therefore, we follow the recommendation of \citet{2014ApJS..211...19B} and synchronize the internal energy using the total energy when deemed safe to do so. We reset $e = (E - \rho|{\bf v}|^2/2)$ if $e/E_{\rm max} \ge 0.1$, where $E_{\rm max}$ is the maximum total energy of the cell and its immediate neighbours [equation (45) of \citet{2014ApJS..211...19B} with $\eta_2 = 0.1$].

\section{Sputtered mass}\label{a:sputtermass}
As each particle is eroded by sputtering, it releases mass into the gas phase. We keep track of the sputtered material by depositing the mass into a passive density field. This field is initially zero and is advected with the flow.

At each time-step, the mass lost by each particle is given by
\begin{equation}
\Delta M_{\rm p} = \frac{4 \pi \rho_{\rm p}}{3}[a^3 - (a - \Delta a)^3],
\end{equation}where $a$ is the current grain radius, $\Delta a = [(da/dt)_{\rm k} + (da/dt)_{\rm t}]~dt$ is the total change in radius due to both non-thermal and thermal sputtering, $M_{\rm p}$ is the mass of each particle, and $\rho_{\rm p}$ is the density of each particle. It is important to note that $\rho_{\rm p} \ne \rho_{\rm d}$, as each particle in \textsc{Athena} represents a collection of many individual dust grains. As the density of each dust grain is fixed at $\rho_{\rm d} = 3.0$ g cm$^{-3}$, the exact number of dust grains per particle depends on the dust mass and the number of particles used; e.g. for 1~M$\odot$ of 1~$\micron$ dust distributed in 10$^5$ particles, each particle represents $\sim 10^{39}$ dust grains. For simplicity, we normalize such that each particle has an initial mass of unity. Hence, if every particle of a given radius group is completely destroyed, the total mass of the passive density field is $N_{\rm p}$. With this simplification, $\rho_{\rm p} = 3/(4 \pi a_0^3)$, where $a_0$ is the original radius of the particle, and 
\begin{equation}
\Delta M_{\rm p} = \frac{a^3 - (a - \Delta a)^3}{a_0^3}.
\end{equation}

\section{Injection efficiencies}\label{a:injection}
The gas injection efficiency is defined as the mass ratio of `injected' gas phase SN ejecta (as traced by the passive colour field $C_{\rm s}$) to the initial amount of gas-phase SN ejecta that is incident on the cloud surface:
\begin{equation}
\eta_{\rm g} \equiv \frac{ \int_V (\rho C_{\rm s})_{\rm injected} }{ \eta_{\rm geom} \int_V (\rho C_{\rm s})_{t=0}}, 
\end{equation}where `injected' material is defined using either the cloud colour field ($C_{\rm c} > 0.1$) or the density ($n > n_{\rm c}$). 

For the dust grain injection efficiency, we must include both sputtered material (traced by the passive density field $\rho_{\rm d}$) and intact grain material. Further, we only consider dust grains that have been stopped, i.e. decelerated to a relative velocity less than 10 per cent of the local sound speed. For each initial radius group, the dust grain injection efficiency is calculated as the mass ratio of both stopped and sputtered material to the initial total particle mass incident on the cloud surface (which we have normalized to be the number of particles $N_{\rm p}$):
\begin{equation}
\eta_{\rm d} \equiv \frac{ [\sum\limits^{N_{\rm p}} (M_{\rm p})_{v_{\rm rel} \le 0.1 c_{\rm s}} + \int_V (\rho_{\rm d})]_{\rm injected} }{ \eta_{\rm geom} \sum\limits^{N_{\rm p}} (M_{\rm p})_{t=0}}.
\end{equation}

%%%%%%%%%%%%%%%%%%%%%%%%%%%%%%%%%%%%%%%%%%%%%%%%%%

% MNRAS
% Don't change these lines
\bsp	% typesetting comment
\label{lastpage}
\end{document}